\documentclass[12pt]{article}

\usepackage{latexsym,amsmath,amssymb,theorem,epsfig}

\newcommand{\mathsym}[1]{{}}
\def\id{\protect{{1 \kern-.28em {\rm l}}}}

\def\be{\begin{eqnarray}}
\def\ee{\end{eqnarray}}

\makeatletter
\renewcommand\section{\@startsection {section}{1}{\z@}%
                                   {-3.5ex \@plus -1ex \@minus -.2ex}%
                                   {2.3ex \@plus.2ex}%
                                   {\normalfont\large\bfseries}}
\renewcommand\subsection{\@startsection{subsection}{2}{\z@}%
                                   {-3.25ex\@plus -1ex \@minus -.2ex}%
                                   {1.5ex \@plus .2ex}%
                                   {\normalfont\normalsize\bfseries}}

\makeatother

\def \foot {\footnote}
\def \bi{\bibitem}

\def \ha {{1 \over 2}}

\def \ci{\cite}

\def \t {\tau}
\def\S{{\mathcal S} }

\def \d {\del}

\def\a{\alpha}

\def \del{\partial}
\def \a {\alpha}

\def\g{\gamma}
\def\s{\sigma}

\def\ov{\over}

\def\l{\lambda}

\def \k {\kappa}
\def\foot{\footnote}

\def \ci {\cite}

\def \d {\partial}

\def \l  {\lambda}

\def \S {{\rm S}}

\def \D {\Delta}

\def \bi{\bibitem}
\def \la {\label}

\def \l {\lambda}
\def\foot{\footnote}

\def \sql {{\sqrt \l}}
\def \adss {$AdS_5 \times S^5~$ }
\newcommand{\rf}[1]{(\ref{#1})}
\def \ov {\over}

\def \ha{{1\ov 2}}
\def \r {\rho}

\def \no {\nonumber}

\def \del {\partial}

\def \S {{\cal S}}

\def \x {{\xi}}

\def \bi{\bibitem}
\def \la {\label}

\def \l {\lambda}
\def\foot{\footnote}

\def \sql {{\sqrt \l}}

\def \adss {$AdS_5 \times S^5$\ }

 \def \t {\tau}
 
 \def \r {\rho}

\def \ov {\over}

\def \varpi {{\rm w}}

\def \pa{\partial}

 \def \t {\tau}

\def \d {\partial}
\def \s {\sigma}

 \def \sql {\sqrt{\lambda}}

\def \Y {{\rm Y}}

\def \S {{\cal S}}

\begin{document}


\overfullrule=0pt
\parskip=2pt
\parindent=12pt
\headheight=0in \headsep=0in \topmargin=0in \oddsidemargin=0in

\vspace{ -3cm}
\thispagestyle{empty}
\vspace{-1cm}

\rightline{Imperial-TP-AT-2010-2}

\begin{center}
\vspace{1cm}
{\Large\bf  

On semiclassical approximation
 for correlators \\ 
 \vspace{0.2cm}
 of closed   string  vertex operators  in AdS/CFT
}
\vspace{1.5cm}
 
 {E.I. Buchbinder\footnote{e.buchbinder@imperial.ac.uk} and A.A.
 Tseytlin\footnote{Also at Lebedev  Institute, Moscow. tseytlin@imperial.ac.uk }}\\

\vskip 0.6cm

{\em 
The Blackett Laboratory, Imperial College,
London SW7 2AZ, U.K.
 }

\vspace{.2cm}

\end{center}

\begin{abstract}
 We consider the  2-point function  of string vertex operators representing 
 string state  with large  spin in $AdS_5$. We compute this correlator in the semiclassical 
 approximation and show that it has the   expected (on the basis of state-operator
 correspondence) form of the strong-coupling limit of the 2-point  function of single trace minimal
 twist operators  in gauge theory. The semiclassical  solution representing the stationary 
 point of
 the path integral   with  two vertex operator insertions   is found to be related to the 
 large spin
 limit of the folded spinning string solution   by a euclidean continuation, transformation to
 Poincare coordinates 
  and  conformal   map from  cylinder to   complex plane. 
 The role of the source  terms coming from   the vertex operator insertions 
   is to specify  the 
  parameters of the solution in terms of  quantum numbers (dimension and spin) of the
    corresponding string state. 
  Understanding further how similar semiclassical methods  may work for 3-point functions  
     may  shed  light  on strong-coupling limit  of the corresponding correlators in gauge 
     theory as
     was recently suggested by Janik et al  in arXiv:1002.4613.

\end{abstract}

\newpage
\setcounter{equation}{0} 
\setcounter{footnote}{0}
\setcounter{section}{0}

\renewcommand{\theequation}{1.\arabic{equation}}
 \setcounter{equation}{0}



\def \ha {{{\textstyle{1 \ov2}}}}
\def \fo {{\textstyle{1 \ov4}}}
\def \C  {{\cal C}}
\def \sql {{\sqrt{\l}}}
\def \S {{\cal S}} 
\def \D  {\Delta }
\newcommand \vev [1] {\langle{#1}\rangle}
\newcommand \VEV [1] {\left\langle{#1}\right\rangle}
\def \ket  {\rangle}
\def \bra  {\langle}
\def \sql  {\sqrt{\l}}
\def \V  { {\rm V}} 
\def \d {\delta}
\def \CC {{\rm C}}
\def \xe  {x_{0e}}
\def \bea {\be} \def \eea {\ee} \def \eqref  {\rf}

\section{Introduction }

The AdS/CFT   duality \ci{mal}  between the  ${\cal N}=4$ SYM theory and the  superstring theory in \adss 
predicts, in particular, a relation between planar correlation functions 
of single-trace  conformal primary operators  in gauge theory  and 
correlation functions of  the corresponding closed-string vertex  operators on 
a 2-sphere. By analogy with the relation  between correlators of BPS operators at strong
coupling  and the supergravity correlators \ci{gkp1,wit} one may   conjecture
the following  equality of the associated generating functionals 
\be 
\bra  e^{  \Phi \cdot O } \ket_{4d} = \bra  e^{ \Phi \cdot  V  } \ket_{2d} \ . \la{ge}
\ee
The l.h.s. in this relation is computed in 4d gauge theory and the r.h.s -- in the 2d superstring 
sigma model. 
 Here  $\Phi(x)$ is a source 4d  field, $O$ is a primary gauge theory operator  
of dimension $\D$  and $V$ is the corresponding marginal string 
vertex operator\foot{Precise relation between 
gauge theory conformal primary operators and marginal vertex operators 
remains an open problem, though  its solution should be aided  by  recent progress in
understanding the spectrum of states based on integrability.}      \ci{pol,tsev} with 
\be \la{de}
&&\Phi \cdot O=  \int d^4 x'\ \Phi(x') O(x') \ ,  \\ 
&&\Phi \cdot  V =\int d^4 x'\ \Phi(x') V(x')=
  \int d^2 \xi \  \hat \Phi (x (\xi),z(\xi) )\  U[x (\xi),z(\xi), ... ] \ ,\no \\
 \ \ \ \ \
&& \hat \Phi (x,z ) = \int d^4 x'  K( x-x'; z  )\  \Phi (x')  \ , \la{ta} \\
&&  V(x')=  \int d^2 \xi \  \V (\xi; x') \ , \ \ \ \ \ \ 
 \V (\xi; x') =  K( x(\xi) -x'; z(\xi)  ) \ U[x (\xi),z(\xi), ... ] \ . \la{jo}
\ee
Here  $\hat x= ( z, x_m)$ are the Poincare patch coordinates,  $ds^2 = z^{-2} ( dz^2 + dx^m
dx_m) $,  and 
\be 
K(  x-x'; z ) 
= c\big[z + z^{-1}  ( x-x')^2\big ]^{-\D}  \ , \ \ \ \ \ \ \ 
K(  x-x'; z  )_{z \to 0} = \delta^{(4)} ( x-x')  \ ,   \la{jp}
\ee
 $U$   stands for  a nontrivial  part of the vertex operator that encodes information about
other quantum numbers.   In general, $U$   depends  also on the   $S^5$ 
coordinates and the fermions present in the superstring action \ci{mt}.\foot{Here we indicated 
 just schematic   form of the  vertex operator. In general, there may be ``mixing''
between the $K$-part and $U$-part  (see \ci{tsev});   this   will not be important in the leading 
semiclassical approximation $\Delta \sim \sql \gg 1$ discussed below.}

The structure of the 2-point and 3-point  correlators in gauge theory  is 
essentially constrained by the 4d conformal  invariance, 
\be 
  \bra O_{\D_1}  (x) O_{\D_2} (x')  \ket_{4d} &=& {C_2\ \delta_{\D_1,\D_2}\ov   |x-x'|^{2 \D_1} }       
 \la{tt}  \ , \\
  \bra O_{\D_1}  (x) O_{\D_2}  (x') O_{\D_3} (x'') \ket_{4d} &=& { C_3   \ov 
  |x-x'|^{\D_1 + \D_2 - \D_3 } |x-x''|^{ \D_1 + \D_3 - \D_2 } 
 |x'-x''|^{\D_2 + \D_3 - \D_1 }}      \la{ttt} \ee
Here $C_2$ depends on normalization of $O_\D (x)$.
 The normalization-independent  combination 
$\bar C_3  \equiv   C_3  [C_2 ({\D_1} ) C_2 (\D_2) C_2 (\D_3)]^{-1/2} $
 is, in general,  a non-trivial function of  `t Hooft coupling $\l$, dimensions 
 $\D_1,\D_2,\D_3$  and other quantum numbers like spins. 
  So far  $C_3$ is explicitly known  for the BPS operators only \ci{Lee,df}. 

On the string side, the world-sheet conformal invariance  and the target space 
$SO(2,4)$ global symmetry of  the string  sigma model  imply that for 
2d  marginal  operators that  represent highest-weight states of $SO(2,4)$ one should get
similar relations (here $J$ stands for some    ``compact'' spin quantum number  that label the
vertex operator)  
\be 
&& \bra \V_{\D_1,J_1} (\xi_1; x)\V_{\D_2,J_2} (\xi_2; x')\ket_{2d}= 
 { \CC_2  \ov   |x-x'|^{2 \D} } \ { \delta(\D_1-\D_2) \delta_{J_1 +J_2,0}  \ov |\x_1 - \x_2|^{2 \delta} }   \ , \la{re}\\
&&   
  \bra V_{\D,J}  (x) V_{\D, -J} (x')  \ket_{2d} = 
  {\C_2\ \ov   |x-x'|^{2 \D} } 
    \ ,  \ \ \ \ \ \ \   \C_2 = \CC_2 \delta(0)  \int {  d^2 \xi_1 d^2 \xi_2  \ov |\x_1 - \x_2|^{2\delta }} \ . \la{ri}
  \ee
  Here  
  \be  \delta (\D, J) =2   \la{dq}  \ee
  is the marginality condition that should be satisfied   for a 
   physical vertex operator  $ \V_{\D,J}$. 

As usual, the definition of  the  string vertex operator  correlators  on a  sphere (complex plane) 
  should  include  
division over the  infinite volume of the  $SL(2,C)$   Mobius  group (which is part of the
 residual  reparametrization gauge 
freedom left out in the conformal gauge). In flat space this   implies the vanishing of
 the  2-point correlators 
 (which has an  interpretation of 
 the  vanishing of the tree-level inverse propagator when evaluated   on the mass shell).
This  should no longer be so in the  $AdS$ case  where the 2-point correlator should 
 have a non-trivial 
  gauge-theory interpretation  \rf{tt}.  The resolution of this puzzle (suggested   
   in the NS-NS   $AdS_3/CFT_2$ 
case \ci{giv,deb} with the  $AdS_3$ part  described by the 
 $SL(2,R)$  WZW model  in \ci{ku,mo})  should  be due to a cancellation between  the divergent factor 
$ \delta(0)  \int {  d^2 \xi_1 d^2 \xi_2  \ov |\x_1 - \x_2|^{4 }}$   and the  Mobius group volume  factor 
contained  in $ \CC_2 $. Here\foot{Note that while $\Delta_i$ take, in general, continuous 
 values, in  the correlator \rf{tt} 
 computed in perturbation  gauge  theory $\delta_{\Delta_1, \Delta_2}$  stands  effectively  
 for the Kronecker delta-symbol  of {\it integer}-valued  
   canonical dimensions of the two operators.} 
the factor  $ \delta(0)=  \delta(\D_1-\D_2)_{\D_1 =\D_2}$
has its origin in the non-compactness of the  symmetry group $SO(1,5)$  of Euclidean $AdS_5$
space which is a global symmetry of the  string sigma model
(this factor  should effectively  come out of the  integral over the  zero mode of the ``Liouville'' 
 field 
$\varphi\equiv \ln z $  where $z$ is the Poincare patch coordinate).\foot{Equivalently, the volume of the  target space conformal symmetries with 2 marked points  should 
 cancel against that of  the  
world-sheet  conformal transformations with 2 marked points. It would be 
useful to  make  this cancellation argument more 
 explicit   by regularizing the divergent factors. 
Let us note also that  the
prescription of \ci{gkp1,wit} for correlators of  the  BPS operators at strong coupling 
  used the  Einstein (or the 
full supergravity) action to define the  correlators on the AdS side. 
The 10d supergravity  action follows from the 
full string theory (expanded near a consistent  vacuum) 
 after one  subtracts 
 Mobius divergences and  massless exchanges  and takes the `` low-energy'' 
 $\a'\to 0$  limit.}

For a 3-point function  of  marginal vertex operators one  should get   
a similar expression as in  \rf{ttt}, i.e.
\be 
&&  \bra V_{\D_1,J_1}  (x) V_{\D_2, J_2}  (x') V_{\D_3,J_3} (x'') \ket_{2d}\no \\ 
 && \ \ \ \ \ \ = { \C_3\ov 
  |x-x'|^{\D_1 + \D_2 - \D_3 } |x-x''|^{ \D_1 + \D_3 - \D_2 } 
 |x'-x''|^{\D_2 + \D_3 - \D_1 }} \ ,    \la{ko}
 \\
 &&
 \C_3 = \CC_3 \int {  d^2 \xi d^2 \xi'  d^2 \xi''
   \ov |\x - \x'|^{2}  |\x - \x''|^{2}    |\x' - \x''|^{2} } \ . \la{ui}
   \ee
Here $\CC_3$  should again contain  the inverse of the  Mobius group factor 
 so that $\C_3$   should be a  finite  function of $\Delta_i, J_i$ and   string tension
and  should  match  $C_3$ in \rf{ttt} (for $AdS_3/CFT_2$ examples see  \ci{mo,else}).

 One may hope to gain some important information  about 
string-theory correlators in semiclassical approximation 
  by considering 
states with large quantum  numbers of order of  string tension $T = { \sql \ov 2 \pi}
 \gg 1 $, i.e. $\D_i \sim  \sql, \  J \sim \sql $. 
That  may then allow one to predict  the strong-coupling behaviour  of 
the corresponding gauge theory
correlators for non-BPS states. 
Indeed, as in the  analogous  limit in  flat space  \ci{gro}, 
if each of the  vertex operators scales  effectively as an  exponent
of the string tension, 
then   the leading $\sql \gg 1$ contribution to the corresponding string path
integral should  be determined by  a classical stationary point. 
In the case of the 2-point functions of vertex operators 
(and related Wilson loop correlators)  such semiclassical approximation 
was  discussed 
 in \ci{pol,gkp2,zar,zap,tsev,yo,tsu,Sak,b1}.\foot{Let
  us mention also a related discussion 
 of semiclassical approximation with vertex operators in $AdS_3$ context in \ci{deb}. 
 In that case  of string sigma model  being $SL(2,R)$  WZW  theory the situation is simpler as the 
 WZW  equations of motion can be solved explicitly.}

Recently, a similar idea was  suggested \ci{ja} for the 3-point correlators. 
While the semiclassical  approximation  should  capture 
the $x$-coordinate dependent factors  in \rf{ko} which scale  as $e^{\sql (...)}$
for $\Delta_i \sim \sql $ (as was, indeed, demonstrated in an  example  in  \ci{ja}) 
it is not   a priori  clear  if it is  sufficient to 
shed light on the strong-coupling  asymptotics of the  non-trivial coefficient 
$\CC_3$, i.e.  whether  it  also scales as $e^{\sql (...)}$ at $\sql \gg 1$
 (such scaling  does not
happen  in the  case of   correlators of three    BPS  operators \ci{Lee}).

The discussion in \ci{ja}
 was qualitative in that it did not specify 
precisely  which are  the string states under consideration, 
 beyond the  fact that  they should  correspond to 
known  classical spinning string solutions on a 2-cylinder.
To be able  to address the question  about strong-coupling asymptotics of correlators 
systematically one needs   to define  the states using vertex operators. 
One should then gain a  better 
understanding of how  a particular structure of the  vertex operators 
representing string states   with large quantum numbers 
translates into the boundary conditions for the corresponding semiclassical world
surface,  apart from a heuristic   requirement that it should end 
at boundary   points in the Poincare patch \ci{wit,ja}  that are  $x$-space locations of
the vertex operators.


\

This will be our aim below.
We shall mainly concentrate on the example of 2-point function 
of vertex operators    representing   spinning string state in $AdS_5$
with $S= \sql \S, \ \S \gg 1$. 
As was  suggested in \ci{tsev}, computing this 
2-point  function  semiclassically  and demanding  the  2d conformal invariance, 
i.e. the consistency of the result  with \rf{ri},\rf{dq},  
should lead to the same relation between the energy (dimension $\D$)
and the spin as found  for the corresponding folded string classical   solution \ci{gkp2}
in the large spin limit. In \ci{tsev} this was shown to be true in the flat space limit;
a  detailed argument for a large spin limit of the spinning string in $AdS_3$ 
(and similar orbiting string states in $S^5$) 
 was recently presented in \ci{b1}. Ref. \ci{b1} used global coordinates 
  while here 
 we shall show how a similar argument can be given in the 
 Poincare coordinates, including also  the   dependence of vertex operators 
 on the boundary points  ignored in \ci{b1}. This   will allow us to  clarify  a  relation 
 of this  vertex  operator approach to the approach of  \ci{ja}. 
 We shall  see    that the 
  corresponding semiclassical world surface  is  closely 
  related  to the euclidean counterpart of the  Poincare  patch image \ci{krtt} of the 
    long folded spinning  string -- the  null cusp surface \ci{kr}.

 We shall review the basic ideas \ci{tsev,b1} of the semiclassical approach  to 
  the 2-point function of the 
spinning string  vertex operators 
  in section 2.  
  In section 3 we shall consider the case when vertex operators represent 
  a string  which is point-like in $AdS_5$  but may be extended in $S^5$. 
  In section 4 we shall discuss our main example --  the semiclassical computation of the 
  2-point function of vertex operators  representing spinning string in $AdS_5$. 
  Some concluding remarks will be made in Section 5. Appendix A  contails details 
  of the argument in section 4 that the stationary point solution for 
  the 2-point correlator is the same as  conformally mapped  euclidean continuation of the 
  large spin limit of the folded string in $AdS_5$.

 \renewcommand{\theequation}{2.\arabic{equation}}
 \setcounter{equation}{0}

 \section{Semiclassical string states and vertex operators}
 
 Let us start with a  review of  the    relation  between 
 semiclassical string states   and  2-point functions of the 
corresponding vertex operators. 
Given  a quantum   closed  string state on a  Minkowski 2d cylinder $(\tau,\s)$
one may first switch  to euclidean time $\tau_e=i \tau$   and 
 map the  cylinder into complex plane with two punctures at  0  ($\tau_e=-\infty$)  and 
$\infty$ ($\tau_e=+\infty$) by $z = e^{\tau_e + i \sigma}$. 
According to  the standard  state-operator correspondence  such  state 
 may be thought of    as created from the vacuum by
a vertex operator inserted at the origin of the  complex plane.\foot{There is a similar
relation between  4d CFT   state on $R \times S^3$ and  the corresponding primary operator
at the origin of $R^4$. The correspondence between a closed string state on $R \times S^1$ 
in global $AdS_5$  and  a 
gauge theory state on $R \times S^3$ thus translates into a  correspondence between  a single trace 
primary operator at the origin of $R^4$   and a  marginal  vertex operator 
 at the origin of $R^2$.} 
 A semiclassical string state with large  quantum numbers of order of string tension 
 can be described by a classical  solution on original Minkowski 2d cylinder.
 The Virasoro condition for the classical solution (that relates its energy to quantum numbers
 like spins)  is then equivalent (for large quantum numbers) 
 to the marginality condition 
 for the vertex operator.\foot{Below we shall use the  string action in the conformal gauge. 
 In general, to recover the two conditions  following from the Virasoro constraints
 one should keep the 2d   metric  arbitrary   and extremise  with respect to 
  it before fixing the conformal
 gauge.}

\subsection{General ideas}
 
When rotating to euclidean world-sheet time it is natural to do similar
 rotation in the target space, i.e.  rotate  the global AdS time $t$ 
\foot{That preserves, in particular,  the reality 
of the space-time energy for a simple   class of solutions. 
In $AdS_5$ case  this allows the classical trajectory  (e.g. massive geodesic) 
to reach the boundary  \ci{yo}.}
\be \la{ro}
\tau_e=i \tau\ , \  \ \ \ \ \ \ \ \ \ \ \   t_e= it \ . \ee
Considering  two vertex operators it is useful to  map the  euclidean 2d cylinder into the 
  complex plane  so that the punctures appear at 
 arbitrary finite  positions $\x_1$ and $\x_2$  
 \be 
\la{pu}
e^{\tau_e + i \sigma} = { \x - \x_2 \ov  \x - \x_1} \ ,  \ee
or, explicitly,  
\be
\tau_e=\frac{1}{2} \ln {(\xi-\xi_2)(\bar \xi -\bar \xi_2) \ov (\xi-\xi_1)(\bar \xi -\bar
\xi_1)}   \ , \ \ \ \ \ \ \ \ \  \s =\frac{1}{2i}  
\ln { (\xi-\xi_2) (\bar \xi -\bar
\xi_1)    \ov  (\bar \xi -\bar \xi_2 )(\xi-\xi_1) }   \ . \label{pul}
\ee
One  may 
  expect that the analytically continued version of a classical  solution 
on 2d cylinder  mapped onto complex plane should then  be  the same as the  world-surface 
which is 
 the stationary trajectory  of the path integral representing the  2-point correlation 
 function of the  vertex operators. This stationary point solution  should solve the 
 rotated (euclidean-signature)  string equations of motion on the complex plane 
 in conformal gauge with  ``delta-function'' sources at $\x_1$ and $\x_2$. 
   The role of matching onto these  source terms
  is to relate the parameters  of the semiclassical  solution 
 to the quantum numbers (energy, spins, ...)  that label  the vertex operators.
The requirement that the resulting correlator evaluated on the stationary point 
solution  takes the conformally invariant form
\rf{re},\rf{dq}  proportional  to  $|\x_1 - \x_2|^{-4}$  is equivalent to the condition 
of  marginality of the vertex operators. It    will   
provide a relation between the parameters which is the same  (for large quantum numbers) 
as the one 
that follows from the  Virasoro  condition for the original solution on the  cylinder.

This equivalence between  the  analytically continued  and conformally mapped 
classical world surface solution  and the  stationary  point  in the corresponding 2-point
 correlator of the
vertex operators  was  explicitly demonstrated in \ci{tsev} in flat space 
on the example of the 
 vertex operator  that describes   bosonic 
 string state on the leading Regge trajectory 
with spin  $S$   and energy $E$  
\be 
\V (\xi) = e^{ -i E t } \big( \pa X\bar{\pa} X \big)^{S/2}\ ,  \  \ \ \ \ \ 
\ \ \ \  X = x_1+ix_2   \   \la{vr} \ee
with  the  marginality  condition being 
$S  + \ha \a'   E^2 =2$, i.e. $E= \sqrt{ 2 (S-2)/\a'}$. 
The relevant  classical solution on  Minkowski 2d cylinder is 
\be \la{cl} 
t = \k \tau \ , \ \ \ \ \
 X= x_1 + i x_2 = \ r(\s) \ e^{ i \phi(\tau)}=\   r_0\  \sin \s\  e^{ i  \tau}   \  , 
\ee
where 
\be \k=r_0 \ , \ \ \ \  \ {\rm i.e.} \ \ \ \ \ \ \ \
   E= \sqrt{ 2 S/\a'} \la{vi} \ee 
     follows from   the  conformal gauge constraint. 
Assuming $S$  is large  and transforming \rf{cl}  using \rf{ro},\rf{pu} one gets a solution 
that indeed solves the string equations with source terms coming from the two vertex operator
insertions, with the $E(S)$ relation  \rf{vi}  now following from the 2d conformal invariance
of the resulting semiclassical value of the  correlator.\foot{The shift   by 2 ($S  + \ha \a'   E^2 =0      \to 
 S  + \ha \a'   E^2 =2  $)
in the marginality condition is of course not visible in the leading  semiclassical large $S$ 
approximation.}

It was suggested  in \ci{tsev} that the same   relation
between the classical solution and the stationary point  world surface in the  vertex operator
correlator 
 should generalise  to the case 
of the spinning folded string solution   in $AdS_5$ \ci{gkp2}. 
The latter   should  correspond 
 to a vertex operator like (cf. \rf{jo})\foot{We mixing with other structures \ci{tsev} and 
  fermionic   contributions  that should be
irrelevant at leading order  of large spin or semiclassical expansion.} 
\be \la{ch} 
\V (\xi) = (Y_5 +  iY_{0})^{-E}\big( \del \Y \bar{\del} \Y \big)^{S/2}\ , 
\ \ \ \ \ \ \ \ \     \Y \equiv  Y_1 + i Y_2  \ .  \ee
Here  $Y_m$ are embedding coordinates of  $AdS_5$ (see below).
By discrete symmetry of the euclidean $AdS_5$  ($i Y_0 = Y_{0e} \leftrightarrow Y_4$)
 this operator  is related to \ci{tsev} 
\be \la{cha} 
\V (\xi) = (Y_5 +  Y_4)^{-\D}\big( \del \Y \bar{\del} \Y \big)^{S/2}\   ,  \ee
with $\D$  being the analog of $E$ in the 54 boost plane. 

The original folded-string solution 
 solves Minkowski  $AdS_5$  string equations 
defined on a   Minkowski 2d cylinder.
It describes propagation of a particular semiclassical  closed  string
mode in real time.  The vertex-operator 2-point
correlator  defined  on a Euclidean 2-plane 
may be mapped onto Euclidean 2-cylinder with the vertex
operators specifying a particular  string state
propagating on the cylinder.
As in flat space case 
the relevant semiclassical trajectory should then  be a (complex
and conformally transformed)
analog of the folded string solution.\foot{As was already apparent  in
the BMN state case with  $\D=J$ case \ci{pol,tsev,yo}, one should not
 attribute a special meaning to the
complex nature of the resulting  semiclassical trajectory
saturating the path integral.
 Its  imaginary  nature of  is  related  to 
 external
sources one puts in to specify the required
 boundary/initial  conditions.
Like in the case of a euclidean gaussian path integral
 with imaginary sources   the result is an analytic function of 
 the quantum numbers,  i.e.   one is allowed to  make analytic continuations.}
 Despite the general expectation that 
 the role of the two  vertex operators  should  simply   be to
implement a  proper choice of boundary conditions in mapping the 2-plane back 
onto the cylinder, i.e.  their  detailed pre-exponential form should not be  that
important, 
 the non-linear nature of the string equations and 
the non-trivial  elliptic function form of the  solution of \ci{dev,gkp2} 
precluded  the direct verification of this relation  in \ci{tsev}.

One  simplification that one can make is to consider the large spin 
 limit of the folded string solution in $AdS_5$ 
in which it  is  stretched all the way to the boundary   and takes a simple (``homogeneous'') 
 form \ci{ftt}.  As was recently shown in \ci{b1}, in this case 
one can  verify that the euclideanized and conformally  transformed \rf{pu} 
large spin solution represents indeed the semiclassical  trajectory saturating the 2-point 
correlation function  of the vertex operators \rf{ch}, with 
the marginality condition being  equivalent to $E= S + {\sql \ov \pi} \ln S\  , \ \ 
{S\ov \sql} \gg 1 $. 

In  \ci{b1} the vertex operator  was chosen  in the  simplest global 
coordinate form  \rf{ch}  with $Y_5+ i Y_0  = \cosh \r\ e^{it}$ \ 
 in which there was no dependence on a boundary point $x_m$. 
This is analogous to ignoring  extra spatial 
 momentum dependence in  the exponent  of the flat-space  
 vertex operator in  \rf{vr}.   As a result, 
the $x$-dependence  of the 2-point correlator in \rf{ri} was not 
determined explicitly. 
   
Motivated by  a  possibility  of  generalization to 3-point correlators 
and in order   to establish the relation to the  approach of \ci{ja} 
(were similar semiclassical computations of 
2-point functions were done without using 
explicit vertex operators)
here  we shall  reconsider the case of the 2-point correlator 
of large spin operators using the Poincare coordinates
and  explicitly including the dependence 
on the boundary points. 
We shall show  that  the large  spin limit of the folded 
string  transformed into the   euclidean  Poincare coordinates 
and  mapped into complex plane using \rf{pu}
represents    the semiclassical surface saturating the  large tension limit of the 
2-point  correlation function \rf{ri}.

\subsection{$AdS_5$ coordinates and vertex operators}

  Let us first recall  the basic  definitions of global and Poincare coordinates 
  in $AdS_5$, their euclidean continuation  and the  form   of the corresponding vertex
  operators \ci{tsev}.
  For  the  standard Minkowski signature  $AdS_5$ we have  
  ($m=0,1,2,   3; \ i=1,2,3$)
\be
&&Y_5 + iY_0 =\cosh \rho\ e^{it}\,, \quad 
Y_1 +i Y_2=\sinh \rho\ \cos\theta\ e^{i \phi_1}\,, \quad
Y_3 +i Y_4=\sinh \rho\ \sin\theta\ e^{i \phi_2}\,,
\nonumber \\
&& Y^M Y_M= -Y_5^2 +Y_m Y^m +Y_4^2 =-1\,, \ \ \ \ \ 
Y_m Y^m= - Y_0^2 + Y_i Y_i   \ ,  
\label{y}
\ee
where $(t,\r, \theta, \phi_1, \phi_2)$ are the global angular coordinates  related 
to the Poincare coordinates by 
\be
Y_m=\frac{x_m}{z}\,, \quad
Y_4=\frac{1}{2 z}(-1+z^2 +x^m x_m) \,, \quad 
Y_5 = \frac{1}{2 z} (1+z^2 +x^m x_m)\,   , 
\label{yy}
\ee
with $  x^m x_m = - x_0^2 + x_i x_i$. 
Assuming a highest-weight state 
of $SO(2,4)$ is labelled by the three  Cartan generators $(E,S_1,S_2)$ 
in the 50,12,34  planes,  
a  wave function or a vertex operator representing a state 
  with  AdS energy  $E$  
should   contain a factor
\be \la{w}
   (Y_5 + iY_0)^{-E} = (\cosh \rho)^{-E} \  e^{-iEt}
	      \ . \ee 
This  does not have a particularly simple form when written in the
Poincare coordinates.	      
If instead we label representations by an $SO(1,1)$ 
 generator in the 54 plane then the corresponding factor  will be 
 \be \la{ww}
  (Y_5 + Y_4)^{-\D} = \left(z+ z^{-1} x^m x_m\right)^{-\Delta}
	      \ ,  \ee 
where  $\D$ is the dilatation generator  ($z\to  k z,\ x_m \to  k x_m$).
	      
The euclidean continuation  corresponds to changing the time-like  coordinates to 
\be \la{p}
t_e=i t  \ , \ \ \ \ \ \    Y_{0e}= i Y_0 \ , \ \ \ \ \ \ x_{0e} = i x_0 \ ,  \ee
so that eq. \rf{y} takes the form  
\be  Y^M Y_M= -Y_5^2 +Y_{0e}^2   + Y_i Y_i  +Y_4^2 =-1\,.  
\label{yu}
\ee
Then the  $SO(2,4)$ symmetry  becomes $SO(1,5)$ which   containes  the discrete  transformation 
\be \la{try}
Y_{0e} \leftrightarrow Y_4 \ , \ \ \ \ \ \ \ \ E \leftrightarrow \D \  \ee
 that relates the factors in   \rf{w}
and \rf{ww}. 

To construct a vertex operator parametrized  by 4 coordinates of a 
 point at the boundary of the  euclidean  Poincare patch 
we should  shift $x_m =(\xe,x_i)$ by a constant vector $x'_m $, getting
the same expression as in \rf{jo},\rf{jp}\foot{Since
translations of $x_m$ are  part of the conformal group 
 this corresponds to a particular  $SO(1,5)$ transformation applied to \rf{ww}.}
\be 
 V(x')=  \int d^2 \xi \ \Big[z(\xi) + z^{-1}(\xi)\ | x_m(\xi) - x'_m|^2\Big]^{-\Delta} 
  \ U[x (\xi),z(\xi), ... ] \ . \la{jk}
\ee
It should be noted that the choice  of the Poincare coordinate form of the vertex operators 
  is  natural for  comparison to the 
standard form of 
correlators of primary  operators on the  gauge theory side (cf. \rf{tt},\rf{ttt}) 
but is not   a priori the only one possible on the string theory side:
one may choose  any  coordinates as long as one uses the 
general form of the vertex  operators that includes dependence 
 on the boundary data.\foot{For example, to construct the analog of \rf{jk} 
  in global coordinates 
one is to  apply the inverse of the 
transformation \rf{yy} to $z(\xi)$ and $x_m (\xi)$ there.}

 \renewcommand{\theequation}{3.\arabic{equation}}
 \setcounter{equation}{0}

\section{Semiclassical approximation: case of point-like string     in $AdS_5$}

Let us  start  the discussion of semiclassical computation of 2-point function of 
operators \rf{jk}  with the case   when $V$ represents 
a state which is point-like in $AdS_5$
(e.g., it may be  a string spinning only  in $S^5$).
Then   $U$  will not depend on $z,x_m$  coordinates.
The simplest  example   is    the BMN state  when 
$U= (X_1 + iX_2)^{-J}$  where $X_I$ are embedding coordinates of $S^5$.\foot{The  euclidean 
stationary point 
 solution for coordinates of $S^5$ will be in general 
 complex, see \ci{pol,zar,yo,tsev,b1}.}
The $AdS_5$ part of the corresponding classical solution 
$t= \kappa \tau$,  $\ \r,\theta,\phi_1,\phi_2=0$ (with $\k$ related to the 
spins of the $S^5$ part via the conformal gauge constraints)
  will be simply a massive geodesic running through the center of $AdS_5$  but   after the euclidean
continuation   it will  be able to reach the boundary  (see  \ci{yo,ja}).  Indeed, 
continued to the Euclid  and  written 
in the  Poincare coordinates \rf{yy} this background  takes the form 
(see, e.g., \ci{tsr})
\be 
z=\frac{1}{\cosh(\kappa \tau_e)}\,, \qquad \xe =\tanh(\kappa \tau_e)\,,\ \ \ \qquad
x_i=0\,,
\label{2.5}
\ee
so that for $\tau_e \to \pm \infty$ we have  $z \to 0$, i.e. the euclidean trajectory 
reaches the boundary at the two points:  $\xe=-1,\ x_i=0$ and  $\xe=1,\ x_i=0$.
 By a dilatation and translation, we can put the position of the end-points
at $\xe=0$ and $\xe=a$;  the corresponding solution is then 
\be
z=  \frac{a}{ 2\cosh(\kappa \tau_e)}\,,  \qquad \xe =\frac{a}{2} \big[
\tanh(\kappa \tau_e)+  1 \big] 
\,, \qquad
x_i=0\,.
\label{2.6}
\ee
Let us now show that  mapping this solution onto the  complex plane by \rf{pu} 
gives a singular configuration  that coincides with the  semiclassical  trajectory 
for the correlator of  two  vertex operators \rf{jk}  inserted 
at the points $x_m=(0,0,0,0)$ and $x'_m = (a,0,0,0)$. 
Assuming $\D \sim \sql \gg 1 $  so that the  correlator  can be
 approximated by a semiclassical
trajectory,  
 the corresponding  euclidean string  action in conformal gauge 
 including the   contributions of the  source terms coming from the
two vertex operator insertions is
($\del  = \ha ( \del_1 + i  \del_2)$) 
\bea
A_e & = & \frac{\sqrt{\lambda}}{\pi} \int d^2 \xi\ \frac{1}{z^2}
(\partial z \bar \partial z + \partial x_m \bar \partial x_m)
- \Delta \int d^2 \xi\ \delta^2(\xi-\xi_1)\ \ln \frac{z}{z^2+ (x_{0e}-a)^2}
\nonumber \\
&  & \ \ \ \ \  \ \ \ \ -\ \Delta \int d^2 \xi\ \delta^2(\xi-\xi_2)\ \ln  \frac{z}{z^2+ x_{0e}^2}
\ + \ A_e(S^5)\,.
\label{2.3}
\eea
Here $A_e(S^5)$ is the $S^5$ part of the action  which 
will be relevant only  for the $|\xi_1-\xi_2|$-dependent 
 part of the correlator, i.e. for the  check of the 
marginality condition  relating $\D$ with  $S^5$ quantum numbers 
 as in the examples discussed in \ci{tsev,b1}.
The resulting equations are solved by $x_i=0$  while 
the equation 
for $x_{0e}$ reads
\be
\partial \frac{\bar \partial \xe}{z^2}
+\bar\partial \frac{\partial \xe}{z^2} =
\frac{2 \pi\Delta}{\sqrt{\lambda}}\Big[ \frac{\xe-a}{z^2+ (\xe-a)^2}
 \ \delta^2(\xi-\xi_1) + 
 \frac{\xe}{z^2+ \xe^2}
\ \delta^2(\xi-\xi_2)\Big]\,.
\label{2.7} 
\ee
Using   \rf{pul}
and  \rf{2.6} we have  $
\frac{\partial \xe}{z^2}=\frac{2 \kappa}{a} \partial \tau_e
$
 and  then  the l.h.s.  of eq.~\rf{2.7} is just
\be
\frac{2\pi \kappa}{a}[\delta^2(\xi-\xi_2)-\delta^2(\xi-\xi_1)]\,.
\label{2.10}
\ee
On the solution~\eqref{2.6} 
we also have $
\frac{\xe}{z^2+ \xe^2}=-\frac{\xe-a}{z^2+ (\xe-a)^2}=\frac{1}{a} $
%
and then    eq.~\eqref{2.7} is satisfied by \rf{2.6}   provided
\be
\kappa=\frac{\Delta}{\sqrt{\lambda}}\,.
\label{2.12}
\ee
It is  straightforward to  check  that the equation for $z$ following from \rf{2.3} 
\be
&&\partial \frac{\bar \partial z}{z^2}
+\bar\partial \frac{\partial z}{z^2} +
\frac{2}{z^3} (\partial z \bar \partial z + \partial \xe \bar \partial \xe)
\nonumber \\
&&=\frac{\pi \Delta}{\sqrt{\lambda}}\ \frac{1}{z} \Big[
\frac{z^2 - (\xe -a)^2}{z^2 +(\xe -a)^2}\ \delta^2(\xi-\xi_1)
+
\frac{z^2 - \xe^2}{z^2 +\xe^2}\ \delta^2(\xi-\xi_2)\Big]\,
\label{2.13}
\ee
is again satisfied by \rf{2.6}  with \rf{2.12}. 
Evaluating the action \rf{2.3} on the classical solution \rf{2.6} mapped according to  \rf{pul} 
we get (using \rf{2.12} and that on the solution $\frac{z}{z^2+(\xe -a)^2}
 =\frac{1}{a}e^{\kappa \tau_e}, \ \
\frac{z}{z^2+\xe^2} =\frac{1}{a}e^{-\kappa \tau_e}$) 
\be
A_e =2 \Delta \ln a -\frac{\Delta^2}{\sqrt{\lambda}} \ln|\xi_1-\xi_2| +
A_e(S^5)\,.
\label{2.16}
\ee
The final expression for the semiclassical  approximation  for the  
2-point correlator  is then  consistent with  \rf{ri} 
\be 
\langle V_{\Delta, J}(a) V_{\Delta, - J}(0)\rangle\  \sim\ 
\frac{1}{a^{2 \Delta}}
\int d^2 \xi_1 d^2 \xi_2\  |\xi_1- \xi_2|^{{\Delta^2\ov \sqrt{\lambda}} - \g(J)}\,,
\label{2.17}
\ee
where $\g(J)= { J^2 \ov \sql} + ...$  stands  for the $S^5$ contribution.
The condition of the world-sheet conformal invariance, i.e. the 
marginality condition \rf{dq}  then relates $\D$  and  $S^5$ spins   ($J, ...$). This 
 will be the same   relation  as between the energy $E$ and the spins  
that follows from the  Virasoro condition  for the original Minkowski-signature 
 real 
solution representing the corresponding semiclassical string state
(see also \ci{b1} for some explicit examples). 

The scaling  $\sim a^{-2 \D}$ of the 2-point  correlator with the boundary point coordinate 
is of course a simple consequence of the conformal invariance of the classical string action 
and of the fact that  the vertex operators  have  definite scaling dimensions: 
it follows from doing  the rescaling $z\to a z, \ \ x_m \to a x_m$  in the path integral. 
The same  will be true in the  spinning string case discussed in the next section.

 \renewcommand{\theequation}{4.\arabic{equation}}
 \setcounter{equation}{0}

\section{Semiclassical approximation: case of folded spinning  string  in $AdS_5$}

Let us now   consider the   semiclassical   approximation for the correlator of two 
vertex operators \rf{cha}   that represent semiclassical string states 
with spin $S \sim \sql \gg 1$. Written in the Poincare coordinates 
(using \rf{yy},\rf{ww},\rf{jk})  the  euclidean vertex operator  labelled by  the boundary 
point $x'_m = (a,0,0,0)$  reads
\be
V_{\Delta, S} (a) =
c  \int d^2 \xi \ \left[z + z^{-1}  (\xe - a)^2\right]^{-\Delta} 
\Big[ \partial  (\frac{r }{z} e^{i \phi} )\ 
\bar \partial (  \frac{r }{z}e^{i \phi} ) \Big]^{S/2} \ , 
\label{3.3}
\ee
where we set (cf. \rf{cl}) 
\be Y_1 + i Y_2 = { x_1 + i x_2  \ov z} = { r \ov z} e^{ i \phi}  \ . \ee 
We  shall consider the correlator 
\be \la{coo} \bra V_{\Delta, S} (a)\  V_{\Delta, -S} (0) \ket \  , \ \ \ \ \ 
 \   V_{\Delta, -S}= ( V_{\Delta, S})^*  \  \ee
in the limit of $ \D \sim S \sim \sql \gg 1 $, with $ \S= { S \ov \sql} $ being large. 
We are going to demonstrate that the semiclassical trajectory that saturates 
the 2-point correlator is  equivalent to the conformally transformed \rf{pu} 
euclidean continuation \rf{ro} of the asymptotic large spin limit \ci{ftt}
of the  spinning folded string solution in $AdS_3$ \ci{dev,gkp2}, i.e. 
\be
t=\kappa \tau\ , \ \  \quad \phi\equiv \phi_1=  \kappa \tau\,,\ \  \quad \rho=\mu \sigma\ , \ \ \ \ \ \ \ \k=\mu \ . 
\label{3.8}
\ee
Here  $\theta=\phi_2=0$ and   $(\tau,\sigma)$ are coordinates of 2-cylinder. 
 The background 
 \rf{3.8}  approximates the elliptic function solution  \ci{gkp2} in the  limit 
 $\k=\mu \gg 1$ on the interval $\s \in[0, {\pi\ov 2}]$; to get the formal periodic solution 
 on $ 0 < \s \leq 2 \pi$ one  needs to combine together   4  stretches  $\rho=\mu \sigma$ 
 of the folded string 
 as in  Figure 1 (see also  \ci{ft2,bec}). 
\begin{figure}[ht]
\centering
\includegraphics[width=100mm]{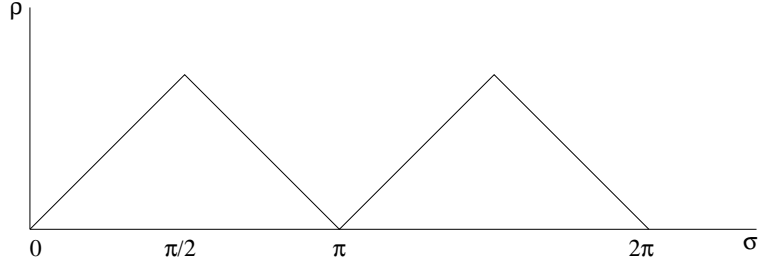}
\caption{The approximate form of  $\rho=\rho(\sigma)$ in the 
limit of large energy and spin. The maximal  value of $\rho$ is $\mu {\pi \ov 2} \gg 1$.}
\label{fig2}
\end{figure}
The condition $\k=\mu$ follows from the Virasoro
constraint and implies  the well-known  relation between the 
corresponding energy and spin \ci{gkp2}
\be
E=S+\frac{\sqrt{\lambda}}{\pi} \ln \frac{S}{\sqrt{\lambda}}+  ... \,,  
\ \ \ \ \ \  \frac{S}{\sqrt{\lambda}}\gg 1   \ .  
\label{3.10}
\ee
Below we shall formally treat  $\k$ and $\mu$ as independent, recovering their equality from the
requirement  of 2d conformal invariance of the semiclassical  value of the 2-point correlator.
 
The formal euclidean continuation \rf{ro} of the solution \rf{3.8} is 
 complex\foot{
Herein an addition to euclidean continuation we (purely for notational convenience) also
 change the  sign of $\phi$ which corresponds to $S \to - S$  and simply interchanges
  the roles of the two operators in the correlator.}
\be
t_e =\kappa \tau_e\,,\qquad \phi = i \kappa \tau_e\,, \qquad \rho=\mu \sigma\,.
\label{3.12}
\ee
Written in the embedding coordinates \rf{yu} it becomes \ci{krtt}
\bea
&&Y_5=\cosh(\k \t_e)\cosh(\mu \s)\,, \qquad Y_{0e}=\sinh(\k \t_e)\cosh(\mu \s)\,,\qquad  Y_4=0 \ ,
\nonumber \\
&&Y_1=\cosh(\k \t_e)\sinh(\mu \s)\,, \qquad Y_2= i \sinh(\k \t_e)\sinh(\mu \s) \ ,  \qquad  Y_3=0 \
.
\label{3.13}
\eea
 In the euclidean  Poincare coordinates \rf{yy}  we then find\foot{For generic
 background $(t,\r,\phi)$  in Minkowski $AdS_3$ coordinates one gets \ci{tsr}:
 $z=  ( \cos  t \ \cosh \r)^{-1} , \ \ x_0 = \tan t, \ x_1 + i x_2 = r e^{i \phi}, \ 
 r = ( \cos t )^{-1}   \tanh  \r$. 
  } 
 \bea && z=\frac{1}{\cosh(\kappa \tau_e) \cosh(\mu \sigma)}\,,\ \ \no \\
&&
 \xe  =\tanh(\kappa \tau_e)\,,\ \  \quad  x_\pm\equiv x_1 \pm  i x_2 = r e^{\pm i \phi} 
 =\frac{\tanh (\mu \s)}{\cosh(\k \t_e)} \   e^{ \mp  \k \tau_e}\, , 
\label{3.14}\\ 
&& z^2 + x_+ x_-  + \xe^2 =1   \ . \la{vv}
\eea
 While in  the Minkowski Poincare
coordinates the string moves towards the center of AdS,
 rotating and stretching,  after the euclidean continuation the world surface 
 described by  \rf{3.14} approaches the boundary ($z \to 0$) 
 at $\tau_e \to \pm \infty$  with $\xe (\pm \infty) = \pm 1$. 
 Performing a simple dilatation and translation 
to ensure that  $\xe (- \infty) = 0$  and  $\xe ( \infty) = a$
as  above in \rf{2.6}   we get 
\bea
&&z=\frac{2}{a}\ \frac{1}{\cosh(\kappa \tau_e) \cosh(\mu \sigma)}\,,\nonumber\\
&& \xe  =\frac{a}{2}\ [\tanh(\kappa \tau_e) +1]\,, 
\ \ \ \ \ \ \ \ 
 x_\pm\equiv x_1 \pm  i x_2 = r e^{\pm i \phi} 
 ={ a \ov 2} \ \frac{\tanh (\mu \s)}{\cosh(\k \t_e)} \   e^{ \mp  \k \tau_e} \ . 
\label{3.16}
\eea
Here the values of the argument $\r= \mu \s$  are  as 
in Figure 1. 
Explicitly, the end-points of the  euclidean world cylinder are  mapped to 
 \be 
&& \tau_e \to  + \infty \ : \ \ \ \   z\to 0 \ , \ \ \xe \to a \ , \ \ r\to 0 \ , \ \ 
 x_+  \to 0 \ , \ \ x_-  \to  {a } \tanh (\mu \s) \ , \la{kp}\\
 && \tau_e \to  - \infty \ : \ \ \ \   z\to 0 \ , \ \ \xe \to 0 \ , \ \ r\to 0 \ , \ \ 
 x_+  \to   {a } \tanh (\mu \s)   \ , \ \ x_-  \to 0 \ . \la{kip}\ee
Note that  if we further analytically continue $x_2 \to i  x_{2t}$ 
to make  the solution real in the Poincare patch  (which will have 
 again the Minkowski signature)  then 
the resulting  surface will be ending  not on points but on null lines.
It will be, in fact,  equivalent  to the null cusp \ci{kr}
surface which was already shown  \ci{krtt} to be   related  by similar transformations\foot{Note
that here  $Y_5$ and $Y_0$ are interchanged compared to \ci{krtt}, i.e. 
to relate the two backgrounds one needs to do   conformal transformations combined with analytic
continuations.} 
to  the  asymptotic form of the folded string solution.

The final step is to  transform the solution \rf{3.16} to the  complex $\xi$-plane
by \rf{pu}, i.e. to  express  $\tau_e$ and $\s$ in \rf{3.16} in terms of $\xi$
using \rf{pul}, with $\tau_e\to \pm \infty$ corresponding to 
$\xi \to \xi_{1,2}$.  As  demonstrated explicitly in Appendix  the resulting solution 
is  the semiclassical trajectory  for the correlator \rf{coo} provided 
the parameters  $\kappa$ and $\mu$ are related 
to the quantum numbers $\D,S$ carried by the vertex operators  as in the 
original folded string solution 
\be
\k =\frac{\Delta-S}{\sqrt{\lambda}}\,, \ \  \qquad \mu =\frac{1}{\pi}
\ln \frac{S}{\sqrt{\lambda}}\,.
\label{3.17}
\ee
The  value of the effective  semiclassical action evaluated on this  solution 
leads   to 
(see Appendix) 
\be
\langle V_{\Delta, S}(a)\ V_{\Delta, -S}(0)\rangle \sim \ 
\frac{1}{a^{2 \Delta}} \ \int d^2 \xi_1 d^2 \xi_2 \  
|\xi_1-\xi_2|^{\sqrt{\lambda}(\k^2-\mu^2)}\,. 
\label{3.18}
\ee
The marginality condition (cf. \rf{ri},\rf{dq})   implies, for large $\k,\mu$,  
\be \kappa \approx \mu\ , \ \ \ \ \ \ \ {\rm i.e. } \ \ \ \ \ \  \ \ 
\Delta \approx S+ \frac{\sqrt{\lambda}}{\pi}\ln \frac{S}{\sqrt{\lambda}}\,.
\label{3.21}
\ee
We thus find that  the 2-point function \rf{3.18}  scales with $a$  as expected 
for a 2-pont function of the  minimal twist gauge theory operators at  strong coupling.


\section{Concluding remarks}

We   have seen how to relate  the   solution representing a  semiclassical string state with large AdS 
spin    with the   stationary point surface in the 2-point correlator of the corresponding vertex
 operators.  Similar relation should  apply to other semiclassical states and  associated vertex operators.
That may help determine the structure of the latter. 

Since classical  closed string solutions  have integrability-based  description 
as finite gap  solutions,  it should be possible to  give an equivalent  description  of 
the stationary  point surfaces ``ending'' at 2 points on  the boundary of the Poincare patch.
It would be interesting to make this description explicit, keeping in mind though that 
generic stationary point surfaces will be {\it complex}. 

An obvious next  question \ci{ja}  concerns  surfaces associated to 3-point correlators and if 
integrability may help  in constructing them. They   should  be related to  classical solutions 
describing  semiclassical decay of a 
spinning string  into two  strings  where one needs to consider the world sheet as a cylinder with a  singular point 
where interaction takes place. 
The resulting surface will  be equivalent to a sphere with 3 punctures.  
Such  solutions may be possible  to construct 
following  \ci{rus,zam}. A related interesting problem is how to 
construct  Wilson loop surfaces ending on  3   finite  contours.

\section*{Acknowledgments}

We  thank  N. Drukker and R. Roiban  for   useful  discussions.
The work of E.I.B is supported by an STFC Fellowship.

\appendix
\section*{Appendix:  Spinning string solution as 
semiclassical trajectory  for 2-point correlator \rf{coo}}

\refstepcounter{section}
\def\theequation{A.\arabic{equation}}
\setcounter{equation}{0}

The Euclidean action including the contribution of  the vertex operators
in \rf{coo}  has the form
\bea
&&A_e = \frac{\sqrt{\lambda}}{\pi} \int d^2 \xi\ \frac{1}{z^2}
(\partial z \bar \partial z +\partial \xe \bar \partial \xe +
\partial r \bar \partial r + r^2 \partial \phi \bar \partial \phi)
\nonumber\\
&& \ - \Delta \int d^2 \xi\ \Big[  \delta^2(\xi-\xi_1) \ln 
\frac{z}{z^2 +r^2 + (\xe - a )^2}  +  \ \delta^2(\xi-\xi_2) \ln 
\frac{z}{z^2 +r^2 + \xe^2} \Big] \nonumber\\
&&\ \ \ -\ \frac{S}{2}\int d^2\xi\ \Big[ \delta^2(\xi-\xi_1) \ln \partial
\frac{r e^{i \phi}}{z}
+ \ \delta^2(\xi-\xi_1) \ln \bar\partial
\frac{r e^{i \phi}}{z}  \nonumber\\
&& \ \ \ \ +\  \delta^2(\xi-\xi_2) \ln \partial
\frac{r e^{-i \phi}}{z}
+  \ \delta^2(\xi-\xi_2) \ln \bar\partial
\frac{r e^{-i \phi}}{z} \Big] \,.
\label{A1}
\eea 
We have set $x_3=0$ as it does not contribute to the semiclassical solution. 
Our aim is to show that the solution to the 
equations of motion following from \rf{A1} is given by \eqref{3.16} transformed to the 
$\xi$-plane using~\eqref{pul}. 

Let us start with the equation for $\phi$. The ``source''  part of it 
will contain terms like
\be
\frac{r e^{i \phi}}{z} \partial \Big(\frac{\delta^2(\xi-\xi_1)}{\partial \frac{r e^{i \phi}}{z}}\Big)
=- \delta^2(\xi-\xi_1)
+\partial\Big(  
\frac{\delta^2(\xi-\xi_1)}{\partial\ln  \frac{r e^{i \phi}}{z}}
\Big)\,.
\label{A3}
\ee
In the limit of large $\mu$,\ $\frac{r e^{i \phi}}{z} \sim e^{\mu \sigma -\k \tau_e}$. 
Then using~\eqref{pul} we find 
\be
\partial\ln  \frac{r e^{i \phi}}{z}  \sim \frac{1}{(\xi-\xi_1)(\xi-\xi_2)}\,, 
\label{A4}
\ee
 and hence
\be
\frac{\delta^2(\xi-\xi_1)}{\partial\ln   \frac{r e^{i \phi}}{z} }=0\,.
\label{A5}
\ee
All delta-function terms can be simplified in the same way. Then 
the equation for $\phi$ becomes
\be
\partial\big (\frac{r^2}{z^2} \bar \partial \phi\big)+
\bar\partial\big(\frac{r^2}{z^2} \partial \phi\big)=
-\frac{i \pi S}{\sqrt{\lambda}}[\delta^2(\xi-\xi_1)-\delta^2(\xi-\xi_2)]\ , 
\label{A6}
\ee
or, using~\eqref{3.16}, 
\be
\partial\big[ \sinh^2(\mu \s) \bar \partial \phi \big]+
\bar\partial\big[\sinh^2(\mu \s) \partial \phi \big]=
-\frac{i \pi S}{\sqrt{\lambda}}[\delta^2(\xi-\xi_1)-\delta^2(\xi-\xi_2)]\,.
\label{A7}
\ee
Exactly the same equation was already discussed  in a slightly different
context in~\cite{b1}. For completeness, we will repeat its   analysis  here.
For  large $\mu$ we can replace $\sinh^2(\mu \s)$
with $e^{2 \mu \s}$. Next, we know that away from singularities
this equation is satisfied and, hence, our aim is understand
how to match the singular contributions. For this we have to recall 
that the actual solution must be a periodic function of $\sigma$, that is
$e^{2 \r} = e^{2 \mu \s}$ should be understood in the sense of Figure 1. 
This  periodic function may  be expanded in Fourier series in $\s\in (0, 2\pi)$, i.e. 
$ f(\s) = \sum_n c_n e^{in \s}$ and then $\s$ may be replaced by $\xi$ using \rf{pul}. 
It is straightforward to see 
 that if $\phi$ has a logarithmic behavior, 
the singular contribution  to  the l.h.s. of \rf{A6} 
 may   only come from the constant Fourier mode of 
%
\be
\left(e^{2 \mu \sigma}\right)_0 =\int_0^{2 \pi}  \frac{d \s}{2 \pi}\ 
e^{2 \mu \s} = 4 \int_0^{\pi/2}  \frac{d \s}{2 \pi}\ 
e^{2 \mu \s} =\frac{1}{\pi \mu}(e^{\pi \mu}-1)\,.
\label{A8}
\ee
If we now substitute~\eqref{A8} into~\eqref{A7} we find the expected solution 
for $\phi$
\be
 \phi =i\kappa\tau_e= \frac{i}{2} \kappa\big(\ln |\xi-\xi_2|^2 -
\ln |\xi-\xi_1|^2\big) 
\label{A9}
\ee
provided $\mu$ is related to $S$ as 
\be
\mu=\frac{1}{\pi}\ln \frac{S}{\sqrt{\lambda}}+\dots\,.
\label{A10}
\ee
%
Now let us  consider the equation for $\xe$
\be
\partial \frac{\bar \partial \xe}{z^2} 
+\bar \partial \frac{\partial \xe}{z^2} =
\frac{2 \pi \Delta}{\sqrt{\lambda}}\Big[\frac{\xe - a }{z^2+r^2 +(\xe - a )^2}\delta^2(\xi-\xi_1)
+\frac{\xe}{z^2+r^2 +\xe^2}\delta^2(\xi-\xi_2)\Big]
\label{A11}
\ee
Using the form of the solution in $(\tau_e, \sigma)$ coordinates~\eqref{3.16}
we can write  this equation as
\be
\k \partial \big[\cosh^2 (\mu \s)\bar\partial \tau_e\big]+
\k \bar\partial \big[\cosh^2 (\mu \s)\partial \tau_e\big]=
\frac{\pi \Delta}{\sqrt{\lambda}}\big[\delta^2(\xi-\xi_2)-\delta^2(\xi-\xi_1)\big]\,.
\label{A12}
\ee
Comparing this  with~\eqref{A7} with 
$\phi$  given in ~\eqref{A9} we conclude  that  eq.~\eqref{A12}
 is satisfied if 
\be
\kappa=\frac{\Delta -S}{\sqrt{\lambda}}\,.
\label{A13}
\ee
Next, let us turn to  the equation for $r$. 
We can simplify the delta-function terms in it  by the
same transformation as in eqs.~\eqref{A3}-\eqref{A5} 
to obtain
\bea
&&r  \partial \frac{\bar \partial r}{z^2}
+ r \bar \partial \frac{\partial r}{z^2}
- 2\frac{r^2}{z^2} \partial \phi \bar \partial \phi
=
-\frac{\pi S}{\sqrt{\lambda}}\big[\delta^2(\xi-\xi_1)+\delta^2(\xi-\xi_2)\big]
\nonumber\\
&&\  + \
\frac{2 \pi \Delta}{\sqrt{\lambda}}\Big[ \frac{r^2}{z^2+r^2 + (\xe - a )^2}
\delta^2(\xi-\xi_1)+
\frac{r^2}{z^2+r^2 + \xe^2}
\delta^2(\xi-\xi_2)\Big]\,.
\label{A14}
\eea
Recalling that $\xi=\xi_1$ corresponds to $\tau_e \to \infty$
and $\xi=\xi_2$ -- to $\tau_e \to -\infty$ we can simplify 
the last two delta-functions as follows
\be
&&\frac{r^2}{z^2+ r^2 +\xe^2}\delta^2(\xi-\xi_2) = 
\tanh^2(\mu \sigma) \ \delta^2(\xi-\xi_2)\ \approx \ \delta^2(\xi-\xi_2) \ , 
\label{A15}\\
&&
\frac{r^2}{z^2+ r^2 +(\xe - a )^2}\delta^2(\xi-\xi_1) = 
\tanh^2(\mu \sigma)\ \delta^2(\xi-\xi_1)\ \approx\ \delta^2(\xi-\xi_1)\,,
\label{A16}
\ee
where we have used that  $\tanh^2(\mu \sigma) \approx 1 $  in the limit 
of large $\mu$. In solving eq.~\eqref{A14} we need to pay attention 
only to  the singular terms.
 Then we obtain (again using  that 
$\mu$ is large)
\be
\kappa \tanh(\kappa \tau_e)\big[
\partial (e^{2 \mu \sigma}\bar \partial \tau_e )+
\bar \partial (e^{2 \mu \sigma}\partial \tau_e )\big]
=-\frac{\pi (2 \Delta-S)}{\sqrt{\lambda}}\big[\delta^2(\xi-\xi_1)+
\delta^2(\xi-\xi_2)\big]\,.
\label{A17}
\ee
Repeating the same analysis as for the equation for $\phi$, using  that
\be \la{A18}
\tanh (\kappa \tau_e)|_{\xi \to \xi_1}=1\ , \  \ \ \ \ \ \ \ \ 
\tanh (\kappa \tau_e)|_{\xi \to \xi_2}=-1\ ,  \ee
and taking into account that to leading order $\Delta \sim S$
we find that eq.~\eqref{A17} is satisfied provided one imposes 
 again the  relation ~\eqref{A10}. 
%

The discussion  of the equation for $z$  is simplified  if 
we  consider the sum of this equation and the 
equation for $r$~\eqref{A14}. Using similar manipulations as above   it 
can be  written in the form
\be
\frac{1}{2} \partial  \frac{\bar \partial (z^2+r^2)}{z^2}
+\frac{1}{2} \bar \partial  \frac{\partial (z^2+r^2)}{z^2}
+\frac{2}{z^2}\partial \xe \bar \partial \xe =
\frac{\pi \Delta}{\sqrt{\lambda}} \big[\delta^2(\xi-\xi_1)+\delta^2(\xi-\xi_2)\big]\,.
\label{A20}
\ee
Using the  solution~\eqref{3.16}   we obtain
\be
\kappa \tanh(\kappa \tau_e)\big[\partial (\cosh^2(\mu \s)\bar \partial\t_e)
+\bar \partial (\cosh^2(\mu \s)\partial\t_e)\big]=
-\frac{\pi \Delta}{\sqrt{\lambda}} \big[\delta^2(\xi-\xi_1)+\delta^2(\xi-\xi_2)\big]\,, 
\label{A21}
\ee
or, in view of ~\eqref{A18}, 
\be
\kappa \big[\partial \big(\cosh^2(\mu \s)\bar \partial\t_e\big)
+\bar \partial \big(\cosh^2(\mu \s)\partial\t_e\big) \big]=
\frac{\pi \Delta}{\sqrt{\lambda}} \big[\delta^2(\xi-\xi_2)-\delta^2(\xi-\xi_1)\big]\,.
\label{A22}
\ee
This is the same as equation \eqref{A12} for $\xe$ which was 
proven to be consistent provided $\kappa$ is related to 
$\Delta$ and $S$ as in~\eqref{A13}. 

This finishes our proof that the solution in~\eqref{3.16} transformed 
 by the conformal map~\eqref{pu} is indeed 
the semiclassical trajectory with singularities prescribed by the 
operators in ~\eqref{3.3},\eqref{coo}. 

Let us now evaluate the action~\eqref{A1} on this solution. 
The string  action (the first line in~\eqref{A1}) is easier 
to compute if we go back to the $(\tau_e, \s)$ cylinder coordinates:
\be
A_{str}=\frac{\sqrt{\lambda}}{4 \pi}(\k^2 +\mu^2)\int d\t_e d \s=
\frac{\sqrt{\lambda}}{2}(\k^2 +\mu^2)(\tau_{e, \infty}-\tau_{e, -\infty})\,,
\label{A23}
\ee
where 
\be
\tau_{e, \pm \infty}=\frac{1}{2}\big(\ln|\xi-\xi_2|^2-
\ln|\xi-\xi_1|^2\big)|_{\xi \to \xi_{1,2}}
\label{A24}
\ee
%
%
%
Ignoring the obvious one-point function divergence ($\sim \ln(0)$) we obtain
\be 
A_{str} =\sqrt{\lambda}(\k^2 +\mu^2)\ln|\xi_1-\xi_2|\, .
\label{A26}
\ee
It is straightforward to evaluate the remaining 
delta-function terms in~\eqref{A1} using the relation 
between $\kappa$, $\Delta$ and $S$~\eqref{A13}; one finds  
\be
2 \Delta \ln a \  - \ 2\sqrt{\lambda}\kappa^2 \ln|\xi_1-\xi_2|\,,
\label{A27}
\ee
where we again  neglected the  $\sim \ln(0)$ terms.
Combining \rf{A26}  and \rf{A27} we end up with  
\be
A_e \approx  2 \Delta \ln a \  -  \sqrt{\lambda}(\k^2 -\mu^2)\ln|\xi_1-\xi_2|
\, ,  
\label{A28}
\ee
so that $e^{-A_e} $  leads to the expression in \rf{3.18}.

\bigskip

\end{document}